\documentstyle[12pt]{article}
\input epsf.sty
\newdimen\psfigsize
\def\psfigure#1 #2 #3 #4 #5{
    \begin{figure}[tbh]
      \begin{center}
      \vbox{
        \null\vskip-0.2in\hskip#2
        \epsfxsize=#1
        \epsfbox{#4}
        \vskip -0.3in
        \caption {#5 \label{#3}}
        \vskip 0.0 true in plus 0.2 true in
      }
      \end{center}
   \end{figure}
}

\def\nab#1{{\nabla_{#1}}}
\def\nabstar#1{\nabla\kern-0.5pt\smash{\raise 4.5pt\hbox{$\ast$}}
               \kern-4.5pt_{#1}}

\def\drvstar#1{\partial\kern-0.5pt\smash{\raise 4.5pt\hbox{$\ast$}}
               \kern-5.0pt_{#1}}

\def\newline{\relax\ifhmode\null\hfil\break\else\nonhmodeerr@\newline\fi}
\def\frac#1#2{{#1\over#2}}
\def\text#1{{\hbox{\rm #1}}}
\def\flushpar{{\par \noindent}}

\newcommand{\beq}{\begin{equation}}
\newcommand{\eeq}{\end{equation}}
\newcommand{\bea}{\begin{eqnarray}}
\newcommand{\eea}{\end{eqnarray}}
\def\Id{ \mbox{1\hspace{-1.2mm}I} }
\begin{document}

\hfill{NTUTH-98-010, April, 1998}

\begin{center}
{\Large{\bf Topological Charge and the Spectrum of \\
            Exactly Massless Fermions on the Lattice}}
\  \\
\  \\
\  \\
{\bf Ting-Wai Chiu } \\
{\normalsize{\em Department of Physics, National Taiwan University \\
   Taipei, Taiwan, R.O.C. \\
}}
\end{center}
\vspace{0.3in}

\begin{abstract}

The square root of the positive definite hermitian operator
$ D_w^{\dagger} D_w $ in Neuberger's proposal of exactly massless quarks
on the lattice is implemented by the recursion formula
$ Y_{k+1} = \frac{1}{2} ( Y_k + D_w^{\dagger} D_w Y_k^{-1} ) $ with
$ Y_0 = \Id $,
where $ Y_k^2 $ converges to $ D_w^{\dagger} D_w $ quadratically.
The spectrum of the lattice Dirac operator for single massless fermion
in two dimensional background $ U(1) $ gauge fields is investigated.
For smooth background gauge fields with non-zero topological charge, the
exact zero modes with definite chirality are reproduced to a very high
precision on a finite lattice and the Index Theorem is satisfied exactly.
The fermionic determinants are also computed and they are in
good agreement with the continuum exact solution.

\end{abstract}

\section{Introduction}

In recent publications \cite{hn97:7,hn98:1}, Neuberger suggested that in a
vector-like gauge theory the lattice Dirac operator $ D $ for exactly
massless quarks can be represented by a finite matrix of fixed shape,
without undesired doubling and no need for any fine tuning.
The lattice Dirac fermion operator of Neuberger's proposal of exactly
massless fermion is
\beq
D_h = \Id + V,   \quad   V = D_w ( D_w^{\dagger} D_w )^{-1/2}
\label{eq:Dh}
\eeq
where $ D_w $ denotes the standard Wilson-Dirac lattice fermion operator
with negative mass term, and $ V $ is unitary ( $ V^{\dagger} = V^{-1} $ ).
$ D_h $ was derived based on the observation that the overlap
\cite{rn95,rn93,rd95} for Dirac fermion in odd dimensions \cite{yk97:7}
can be written as the determinant of a finite matrix of fixed shape.
The most remarkable feature of $ D_h $ is that it satisfies the
Ginsparg-Wilson relation \cite{gwr}
\beq
D \gamma_5 + \gamma_5 D = D \gamma_5 D
\label{eq:gwr}
\eeq
and $ D_h $ is so far the only known $ explicit $ solution of
eq.(\ref{eq:gwr}). Ginsparg-Wilson relation was derived in 1981 as the
remnant of chiral symmetry on the lattice after blocking a chirally
symmetric theory with a chirality breaking local renormalization group
transformation. The original Ginsparg-Wilson relation is in fact more
general than eq.(\ref{eq:gwr}) and constitutes a matrix $ R $
which is local in the position space but diagonal in the Dirac space
\beq
D \gamma_5 + \gamma_5 D = D \gamma_5 R D
\label{eq:gwo}
\eeq
It serves as $the$ criterion for breaking the continuum chiral symmetry
on the lattice while preserving the exact masslessness and the continuum
axial anomaly. Recently, Hasenfratz, Laliena and Niedermayer \cite{ph98:1},
and L\"uscher \cite{ml98:2} explicitly showed that any $ D $ satisfying the
Ginsparg-Wilson relation plus some reasonable assumptions such as locality
and free of species doubling must obey the Index Theorem on the lattice.
Furthermore, L\"uscher \cite{ml98:2} discovered that any $ D $ satisfying
Ginsparg-Wilson relation implies an exact symmetry of the fermion
action which may be regarded as a lattice form of chiral symmetry,
reproduces the correct anomaly and the Index Theorem on the lattice.
The Nielsen-Ninomiya theorem \cite{no-go}
can be circumvented if the continuum chiral symmetry of the fermion is
replaced by the Ginsparg-Wilson relation which is the chiral symmetry
realized on the lattice. Recently, Narayanan \cite{rn98:2} has shown that the
fermionic determinant of Dirac fermion operator $ D $ in the form
$ D = \Id + V $ and satisfying the Ginsparg-Wilson relation can be factorized
into two factors which are complex conjugate of each other, corresponding to
those of left-handed and right-handed Weyl fermions. This could imply
that the dynamical gauge theory of $ single $ massless Dirac fermion using
$ D_h $ is amenable to Hybrid Monte Carlo simulations.
However, the most challenging part of Neuberger's
proposal is the implementation of the square root of the positive defintite
hermitian operator $ D_w^{\dagger} D_w $. The approximation of $ D_h $ has
been studied by Neuberger \cite{hn97:10}, but it sacrifices strictly
masslessness. In this paper, we attempt to implement the square root operation
on the {\em positive definite hermition } operator
$ D_w^{\dagger} D_w $ by the following recursion formula \cite{matrix:83}
\beq
Y_{k+1} = \frac{1}{2} ( Y_k +  D_w^{\dagger} D_w Y_k^{-1} ), \quad\quad
Y_0 = \Id
\label{eq:sqrt}
\eeq
It can be shown that $ Y_k^2 $ converges to $ D_w^{\dagger} D_w $ quadratically.
The main purpose of this paper is to investigate the spectrum of $ D_h $ to
see to what extent this implementation can reproduce the exact zero modes
with definite chirality as well as the realization of the Index Theorem on
a finite lattice. The fermionic determinants are also computed and compared
with the continuum exact solutions.

\section{Massless Fermion Action}

Neuberger's lattice fermion operator for exactly massless fermion has been
given in eq.(\ref{eq:Dh}). The negative mass term in the standard
Wilson-Dirac lattice fermion operator $ D_w $ can be chosen
to be any value in the range $ ( -1, 0 ) $.  A different value of
$ m $ corresponds to a different renormalization for the observables.
In the following, we shall restrict our discussions to the case of
$ m = -1 $. Then the Wilson-Dirac operator becomes
\bea
D_w = - \Id
      + \frac{1}{2} [ \gamma_{\mu} ( \nabstar{\mu} + \nab{\mu} ) -
                      \nabstar{\mu} \nab{\mu} ]
\eea
where $ \nab{\mu} $ and $ \nabstar{\mu} $ are the forward and
backward difference operators defined in the following,
\bea
\nab{\mu}\psi(x) &=& \
     U_\mu(x)\psi(x+\hat{\mu})-\psi(x)  \nonumber \\
\nabstar{\mu} \psi(x) &=& \ \psi(x) -
     U_\mu^{\dagger}(x-\hat{\mu}) \psi(x-\hat{\mu})  \nonumber
\eea
Then the lattice action of single massless fermion in background gauge
field is
\beq
A_h = \sum_{x} \sum_{y} \bar{\psi}(x) D_h (x,y) \psi(y)
\label{eq:Ah}
\eeq
where the Dirac indices are suppressed.

On a torus ( $ x_{\mu} \in [0,L_{\mu}], \mu = 1,2 $ ), the $ U(1) $ gauge
fields can be decomposed into global, harmonic and local parts. In this
paper we use the following decomposition.
\beq
A_1(x) = - \frac{ 2 \pi Q x_2 }{ L_1 L_2 }
         + \frac{ 2 \pi h_1 }{L_1}
         +  A_1^{(0)} \sin ( \frac{ 2 \pi n_2 }{L_2} x_2 )
\label{eq:A1}
\eeq
\beq
A_2(x) = \frac{ 2 \pi h_2 }{L_2}
         +  A_2^{(0)} \sin ( \frac{ 2 \pi n_1 }{L_1} x_1 )
\label{eq:A2}
\eeq
where the global part is characterized by the topological charge
\beq
Q = \frac{1}{2\pi} \int d^2 x F_{12}
\label{eq:ntop}
\eeq
which must be an integer. The harmonic parts are parameterized by two
constants $ h_1 $ and $ h_2 $. The local parts are chosen to be sinusoidal
fluctuatations with amplitudes $ A_1^{(0)} $ and $ A_2^{(0)} $ and
frequencies $ \frac{ 2 \pi n_2 }{L_2} $ and $ \frac{ 2 \pi n_1 }{L_1} $ where
$ n_1 $ and $ n_2 $ are integers. The discontinuity of $ A_1(x) $ at
$ x_2 = L_2 $ due to the global part only amounts to a gauge transformation.
The field strength $ F_{12} = \partial_1 A_2 - \partial_2 A_1 $ is continuous
on the torus.
To transcribe the continuum gauge fields to the lattice, we take the lattice
sites at $ x_\mu = 0, a, ..., ( N_\mu - 1 ) a $, where $ a $ is the lattice
spacing and $ L_\mu = N_\mu a $ is the lattice size.
Then the link variables are
\bea
U_1(x) &=& \exp \bigl[ \text{i} A_1(x) a \bigr]  \\
\label{eq:U1}
U_2(x) &=& \exp \bigl[ \text{i} A_2(x) a
      + \text{i} \frac{ 2 \pi Q x_1 }{ L_1 } \delta_{x_2,(N_2 - 1)a} \bigr]
\label{eq:U2}
\eea
The last term in the exponent of $ U_2(x) $ is included to ensure that the
field strength $ F_{12} $ which is defined by the ordered product of link
variables around a plaquette is continuous on the torus.

The fermion propagator $ S_F(x,y) $ is defined by
\begin{eqnarray}
S_F(x,y) = \frac{1}{Z} \int \prod_z d \bar{\psi}(z) d \psi(z)
           \ \text{e}^{-A_h} \ \psi(x) \bar{\psi}(y)
\end{eqnarray}
where
\beq
Z = \int \prod_z d \bar{\psi}(z) d\psi(z) \text{e}^{-A_h}
\eeq
In background gauge fields of zero topological charge ( $ Q = 0 $ ),
the fermion propagator is
\begin{eqnarray}
S_F(x,y) = D_h^{-1}(x,y)
\label{eq:SF}
\end{eqnarray}
The free fermion propagator in momentum space is
\beq
\tilde{S_F}^{(0)}(p) = \frac{a}{2} \ \Id -
    \text{i} a \ \frac{ \sum_\mu \gamma_\mu \sin(p_\mu a )}{2[N(p)+u(p)]}
\label{eq:SF0}
\eeq
where $ a $ is the lattice spacing and
\bea
u(p) = 1 - \sum_\mu \cos( p_\mu a )              \\
N(p) = \sqrt{ u^2(p) + \sum_\mu \sin^2(p_\mu a) }
\eea
The constant term $ \frac{a}{2} \Id $ which vanishes in the continuum limit
is expected to appear in eq.(\ref{eq:SF0}) such that $ D_h^{-1} $ satisfies the
Ginsparg-Wilson relation
\beq
\gamma_5 D^{-1} + D^{-1} \gamma_5 = a \gamma_5
\label{eq:gwi}
\eeq
which is equivalent to Eq.(\ref{eq:gwr}).
The denominator $ 2 [ N(p) + u(p) ] $ in eq.(\ref{eq:SF0}) has
only one zero at $ p_1 = p_2 = 0 $ for the entire Brillouin zone and
its expansion around $ p_1 = p_2 = 0 $ is
\beq
2[ N(p) + u(p) ] = a^2 ( p_1^2 + p_2^2 ) + O(a^4 p^4 )
\eeq
Therefore the free fermion propagator is free of doublers and has the
correct continuum limit.
We remark that {\em it is trivial to construct $ D^{-1} $ satisfying
Ginsparg-Wilson relation.} Any $ D^{-1} $ in the form
\beq
D^{-1} ( x, y ) = \frac{a}{2} \Id + \sum_\mu \gamma_\mu S_\mu ( x, y )
\label{eq:Dinv}
\eeq
must satisfy eq.(\ref{eq:gwi}). However, {\em the additional requirements
that the resulting lattice fermion action $ ( D^{-1} )^{-1} $
is local, free of species doubling and satisfying the Index Theorem would
make the task rather non-trivial. }
So far, $ D_h $ is the only known {\em explicit } solution fulfilling all
these requirements.

\section{Zero Modes and The Index Theorem}

In the continuum the Dirac operator of massless fermions in a smooth
background gauge field with non-zero topological charge $ Q $ has zero
eigenvalues and the corresponding eigenfunctions are chiral. The Index
Theorem \cite{index} asserts that the difference of the number of
left-handed and right-handed zero modes is equal to the topological charge of
the gauge field configuration :
\bea
 n_{-} - n_{+} = Q
\label{eq:index}
\eea
In two dimensions, the so called Vanishing Theorem \cite{vanish} also holds :
\bea
 Q > 0 & \Rightarrow & n_{+} = 0 \quad \text{ and } \quad Q = n_{-} \nonumber \\
 Q < 0 & \Rightarrow & n_{-} = 0 \quad \text{ and } \quad Q = - n_{+}
\label{eq:vanish}
\eea
Recently, Hasenfratz, Laliena and Niedermayer \cite{ph98:1}, and
and L\"uscher \cite{ml98:2} explicitly showed that any lattice Dirac
fermion operator $ D $ satisfying the
Ginsparg-Wilson relation eq.(\ref{eq:gwr}) must obey the Index
Theorem exactly. Since $ D_h $ is an explicit solution of Ginsparg-Wilson
relation, it must obey the Index Theorem exactly. However,
on a finite lattice, after we implement the inverse square root operator
in $ D_h $ by the recursion formula eq.(\ref{eq:sqrt}), it is not known
{ \em a priori } how the zero modes and the Index Theorem can be recovered.
It turns out that even on a very small lattice ( 6 x 6 ),
the exact zero modes with definite chirality are reproduced to a very
high accuracy and the Index Theorem is satisfied exactly.
For smooth background guage fields, the convergence
of the recursion formula eq.(\ref{eq:sqrt}) is indeed quite fast.
If we require the absolute error of each matrix element satisfying the
criterion
\bea
 |Y^2_{ij} - ( D_w^{\dagger} D_w )_{ij}| < \epsilon
\label{eq:epsilon}
\eea
then it usually takes less than 5 iterations to reach $ \epsilon = 10^{-8} $
for smooth gauge configurations. For very rough configurations, the convergence
could be slow and the accumulation of roundoff errors in successive iterations
could cause the algorithm unstable.
For exceptional configurations,
$ det( D_w )= 0 $, then the inverse square root operator in $ D_h $ is
not defined. This is consistent with the fact that eq.(\ref{eq:sqrt}) breaks
down when $ D_w^{\dagger} D_w $ is not positive definite.
In fact, the convergence of eq.(\ref{eq:sqrt}) has become very poor in the
vicinities of exceptional configurations.

After the matrix $ D_h $ is computed, we solve the following eigenproblem
\bea
  \sum_{y} \sum_{\beta} D_h^{\alpha \beta}(x,y) \phi_s^{\beta}(y) =
  \lambda_s \phi_s^{\alpha} (x) \quad s = 1, \cdots, 2 N_1 N_2
\eea
where $ x $ and $ y $ are site indices, $ \alpha $ and $ \beta $ are Dirac
indices, and $ \phi_s $ is normalized eigenfunction
\beq
 \sum_x \sum_{\alpha} [\phi_s^{\alpha}(x)]^{*} \phi_s^{\alpha}(x) = 1
\eeq
In matrix notations, these equations are rewritten as
\bea
\label{eq:eigenD}
   D_h \phi_s &=& \lambda_s \phi_s  \\
   \phi_s^{\dagger} \phi_s &=& 1
\eea
Before we solve for the eigenvalues and eigenfunctions
of $ D_h $, we discuss some of their general analytical properties
in the following.
Since $ \gamma_5 D_w \gamma_5 = D_w^{\dagger} $,
it follows that
\beq
 \gamma_5 D_h \gamma_5 = D_h^{\dagger}
\label{eq:adjoint}
\eeq
and the secular equation
\beq
  det( D_h - \lambda \Id ) = det( D_h^{\dagger} - \lambda^{*} \Id )
= det [ \gamma_5 ( D_h - \lambda^{*} \Id ) \gamma_5 ]
= det( D_h - \lambda^{*} \Id ) = 0
\eeq
implies that {\em the eigenvalues $ \lambda_s $ are either real or come in
complex conjugate pairs}.  Using eq.(\ref{eq:adjoint}) and
eq.(\ref{eq:gwr}), we obtain
\beq
\label{eq:normal}
  D_h^{\dagger} D_h = D_h D_h^{\dagger} \Longleftrightarrow
  D_h \ {\text{ is normal }}
\eeq
and
\beq
\label{eq:gwh}
  D_h^{\dagger} + D_h = D_h^{\dagger} D_h
\eeq
Since $ D_h $ is normal, $ D_h $ and $ D_h^{\dagger} $ have common
eigenfunction and their eigenvalues come in complex conjugate pairs,
\beq
   D_h^{\dagger} \phi_s = \lambda_s^{*} \phi_s
\label{eq:dhd}
\eeq
and the eigenvectors $ \{ \phi_s \} $ form a complete orthonormal set.
Eq.(\ref{eq:gwh}) implies that
\beq
\lambda_s^{*} + \lambda_s = \lambda_s^{*} \lambda_s \Rightarrow
| \lambda_s - 1 |^2 = 1
\label{eq:circle}
\eeq
Thus the eigenvalues of $ D_h $ fall on a unit circle with center at 1 and
have the reflection symmetry with respect to the real axis.
We define the chirality of an eigenmode to be
\beq
\chi_s = \phi_s^{\dagger} \gamma_5 \phi_s =
\sum_x \sum_{\alpha} \sum_{\beta}
     [ \phi_s^{\alpha}(x) ]^{*} \gamma_5 ^{\alpha \beta} \phi_s^{\beta} (x)
\label{eq:chi}
\eeq
Using eq. (\ref{eq:adjoint}) and eq. (\ref{eq:dhd}), we obtain
\beq
   D_h  \gamma_5 \phi_s  = \lambda_s^{*} \gamma_5 \phi_s
\label{eq:d5p}
\eeq
Multiplying $ \phi_s^{\dagger} $ on both sides and using the eigenvalue
equation, we get
\beq
\lambda_s \phi_s^{\dagger} \gamma_5 \phi_s =
\lambda_s^{*} \phi_s^{\dagger} \gamma_5 \phi_s
\eeq
Then
\beq
 \chi_s = \phi_s^{\dagger} \gamma_5 \phi_s = 0 \quad \text{if} \
 \lambda_s \ne \lambda_s^{*}
\label{eq:chic}
\eeq
{\em The chirality of any complex eigenmode is zero. }
If $ \lambda_s $ is real ( 0 or 2 ),  eq.(\ref{eq:d5p}) and
eq.(\ref{eq:eigenD}) imply that $ \phi_s $ has definite chirality
$ +1 $ or $ -1 $.
\beq
\gamma_5 \phi_s = \pm \phi_s \quad \text{if} \ \lambda_s = \lambda_s^{*}
\label{eq:chiral}
\eeq
It is remarkable that the zero modes and the +2 eigenmodes
are both chiral. This is true for any $ D $ satisfying the Ginsparg-Wilson
relation eq. (\ref{eq:gwr}) and the adjoint condition eq. (\ref{eq:adjoint}).
Another useful property of chirality is that {\em total chirality
of all eigenmodes must vanish },
\bea
\sum_s \chi_s &=& \sum_s \phi_s^{\dagger} \gamma_5 \phi_s  \nonumber \\
&=& \sum_s \sum_x \sum_{\alpha} \sum_{\beta}
[\phi_s^{\alpha}(x)]^{*} \gamma_5^{\alpha \beta} \phi_s^{\beta}(x) \nonumber \\
&=& \sum_x \sum_{\alpha} \sum_{\beta} \gamma_5^{\alpha \beta}
    \delta_{\alpha \beta} = 0
\label{eq:chisum}
\eea
where the completeness relation 
\beq
\sum_s [\phi_s^{\alpha}(x)]^{*} \phi_s^{\beta}(y)
= \delta^{\alpha \beta} \delta_{x y}
\eeq
for the eigenfunctions of $ D_h $ has been used.
From eq. (\ref{eq:chisum}), we immediately obtain
\bea
 n_2^{+} - n_2^{-} = n_{-} - n_{+}
\label{eq:chir}
\eea
{\em The sum of chirality of all +2 eigenmodes is
equal to the index of the zero modes }. Therefore we can identify
the chirality of +2 eigenmodes as the index of the zero modes for
any $ D $ satisfying the Ginsparg-Wilson relation and the adjoint condition.
On the other hand, the standard Wilson-Dirac lattice fermion operator
$ D_w $ satisfies the adjoint condition but not the Ginsparg-Wilson relation.
The chirality of its complex eigenmodes is zero but
{\em its real eigenmodes do not have definite chirality. }
Moreover, it does not have exact zero modes even in a smooth background
gauge field with non-zero topologically charge. Unlike those $ D $ satisfying
Ginsparg-Wilson relation has real eigenvalues only at 0 or 2, $ D_w $ has
real eigenvalues at several different values and the total chirality of all
( real ) eigenmodes does not vanish. Therefore, strictly speaking,
we should not identify the chirality of a subset of real eigenmodes
of $ D_w $ to be the index of zero modes. Recently, Gattringer, Hip and Lang
\cite{gatt97} conjectured that the sum of the chirality of the
real eigenmodes of the Wilson-Dirac hopping matrix in the vicinity of +2D
( D is the dimensionality ) is equal to the minus of the topological charge
( $ -Q $ ) of the background gauge field. However, the Wilson-Dirac fermion
operator does not have exact zero modes without tuning the mass parameter
for each gauge configuration. {\em Only in the infinite
volume limit}, the +2D modes of Wilson-Dirac hopping matrix are equivalent
to the zero modes of the Wilson-Dirac fermion operator with zero bare mass,
it implies that {\em in the infinite volume limit } the approximate zero
modes of the Wilson-Dirac operator will become exactly massless with
definite chiralities and the Index Theorem can be satisfied exactly.

In the following we derive some basic properties of the unitary matrix
$ V = D_h - \Id $ which has eigenvalues $ \lambda_s - 1 $. The unitarity
of $ V $ implies that its eigenvalues are unimodular
( i.e., $ e^{ \text{i} \theta_s } $ ) and the eigenvalues of $ D_h $
are in the form $ 1 + e^{\text{i} \theta_s } $.
The zero modes of $ D_h $ correspond to those of eigenvalues -1 of $ V $.
Since the eigenvalues of V are either real ( +1 or -1 ) or come in
complex conjugate pairs, it follows that
\beq
det(V)= (-1)^{ ( n_{-} + n_{+} ) } = (-1)^{ ( n_{-} - n_{+} ) }
= (-1)^{ ( n_2^{+} - n_2^{-} ) }
\label{eq:detV}
\eeq
where eq.(\ref{eq:chir}) has been used in the second equality.
Eq.(\ref{eq:detV}) can serve as a check for the consistency of
the eigenvalues.

In Table 1, the eigenvalues of $ D_h $ are listed for $ 6 \times 6 $
lattice in the background gauge field [ eqs.(\ref{eq:A1}) and (\ref{eq:A2}) ]
with topological charge $ Q = 1 $; and
harmonic parts with $ h_1 = 0.1 $ and $ h_2 = 0.2 $, and the local
parts with $ A_1^{(0)} = 0.3 $, $ A_2^{(0)} = 0.4 $ and $ n_1 = n_2 = 1 $.
It is evident that the eigenvalues are either real ( 0 or 2 ),
or come in complex conjugate pairs in a very precise manner.
This property is vital to obtain real and positive fermionic determinant.
There is one zero mode with exactly zero eigenvalue and chirality $ -1 $.
Therefore we have $ n_{-} = 1 $ and $ n_{+} = 0 $ and the Index Theorem
$ Q = n_{-} - n_{+} $ is satisfied exactly. Moreover, the Vanishing Theorem
$ Q > 0 \Longleftrightarrow n_{+}=0 $ which only holds in two dimensions is
also satisfied. We also note that eq.(\ref{eq:chisum}) and eq.(\ref{eq:chir})
are both satisfied. All analytical properties of the eigensystem
we have discussed above are satisfied exactly.
The eigenvalues in Table 1 are plotted in Fig. 1.

In Table 2, the topological charge $ Q = - 2 $
while other parameters are the same as in Table 1.
There are two exact zero modes of chirality $ +1 $.
Therefore we have $ n_{-} = 0 $ and $ n_{+} = 2 $ and the
Index Theorem $ Q = n_{-} - n_{+} $ and the Vanishing Theorem
$ Q < 0 \Longleftrightarrow n_{-} = 0 $ are satisfied.
Again all analytical properties are satisfied exactly.
The eigenvalues in Table 2 are plotted in Fig. 2.

We have tested many different smooth gauge configurations by changing the
topological charge as well as other parameters in
eqs. ( \ref{eq:A1} ) and ( \ref{eq:A2} ). The exact zero modes
are always reproduced such that the Index Theorem and the Vanishing
Theorem are satisfied. All analytical properties we have discussed above
are satisfied exactly. We summarize some of our results in Table 3.

We also investigated the robustness of zero modes and the index theorem
under local fluctuations by varying the amplitudes
$ A_1^{(0)}$ and $ A_2^{(0)} $ as well as the frequencies
$ \frac{ 2 \pi n_1}{L_1} $ and $ \frac{ 2 \pi n_2 }{L_2} $
of the background gauge field. We found that the exact zero
modes are very robust under variations of the background. For example,
for $ Q = 1 $ and $ n_1 = n_2 = 1 $, the exact zero modes and the Index Theorem
are reproduced for $ A_1^{(0)}$ and $ A_2^{(0)} $ ranging from 0.0 to 0.9.

The stability of zero modes under random fluctuations are also investigated.
The rough gauge configurations are obtained from the smooth ones by multiplying
each link variable with a random phase \cite{smit87}
\bea
U_\mu(x)_r = \text{e}^{ \text{i} r \theta_\mu(x) } U_\mu(x)
\eea
where $ \theta_\mu(x) $ is a uniformly distributed random variable in
$ ( - \pi, \pi ) $ and $ r $ controls the size of roughness. We found that
for gauge configurations with topological charge $ | Q | \le 6 $ and
$ A_1^{(0)} = 0.3 $, $ A_2^{(0)} = 0.4 $ and $ n_1 = n_2 = 1 $,
the exact zero modes satisfying the Index Theorem can be reproduced
provided that $ | r | \le 0.3 $.

{\footnotesize
\begin{table}
\begin{center}
\begin{tabular}{|c|c|c|c|c|c|}
\hline
Re($\lambda$) & Im($\lambda$) & chirality &
Re($\lambda$) & Im($\lambda$) & chirality \\
\hline
\hline
 2.00000000 &   .00000000 &   1.0000000 & & & \\
 \hline
 1.99981020 &   .01948230 &   .00000000 &  1.99981020 &  -.01948230 &   .00000000 \\
 \hline
 1.98951740 &   .14441370 &   .00000000 &  1.98951740 &  -.14441370 &   .00000000 \\
 \hline
 1.96345153 &   .26788272 &   .00000000 &  1.96345153 &  -.26788272 &   .00000000 \\
 \hline
 1.93881783 &   .34441410 &   .00000000 &  1.93881783 &  -.34441410 &   .00000000 \\
 \hline
 1.93303732 &   .35977961 &   .00000000 &  1.93303732 &  -.35977961 &   .00000000 \\
 \hline
 1.93041388 &   .36651060 &   .00000000 &  1.93041388 &  -.36651060 &   .00000000 \\
 \hline
 1.91625509 &   .40059533 &   .00000000 &  1.91625509 &  -.40059533 &   .00000000 \\
 \hline
 1.90283899 &   .42997879 &   .00000000 &  1.90283899 &  -.42997879 &   .00000000 \\
 \hline
 1.89600810 &   .44403770 &   .00000000 &  1.89600810 &  -.44403770 &   .00000000 \\
 \hline
 1.87536396 &   .48346451 &   .00000000 &  1.87536396 &  -.48346451 &   .00000000 \\
 \hline
 1.86076634 &   .50900030 &   .00000000 &  1.86076634 &  -.50900030 &   .00000000 \\
 \hline
 1.84065148 &   .54157648 &   .00000000 &  1.84065148 &  -.54157648 &   .00000000 \\
 \hline
 1.83027494 &   .55735405 &   .00000000 &  1.83027494 &  -.55735405 &   .00000000 \\
 \hline
 1.80483165 &   .59350317 &   .00000000 &  1.80483165 &  -.59350317 &   .00000000 \\
 \hline
 1.78430700 &   .62037289 &   .00000000 &  1.78430700 &  -.62037289 &   .00000000 \\
 \hline
 1.76077457 &   .64901623 &   .00000000 &  1.76077457 &  -.64901623 &   .00000000 \\
 \hline
 1.73598092 &   .67700228 &   .00000000 &  1.73598092 &  -.67700228 &   .00000000 \\
 \hline
 1.71413711 &   .70000585 &   .00000000 &  1.71413711 &  -.70000585 &   .00000000 \\
 \hline
 1.68715242 &   .72651329 &   .00000000 &  1.68715242 &  -.72651329 &   .00000000 \\
 \hline
 1.65416649 &   .75635058 &   .00000000 &  1.65416649 &  -.75635058 &   .00000000 \\
 \hline
 1.61565168 &   .78801841 &   .00000000 &  1.61565168 &  -.78801841 &   .00000000 \\
 \hline
 1.57389192 &   .81893105 &   .00000000 &  1.57389192 &  -.81893105 &   .00000000 \\
 \hline
 1.51607964 &   .85654060 &   .00000000 &  1.51607964 &  -.85654060 &   .00000000 \\
 \hline
 1.47365734 &   .88070922 &   .00000000 &  1.47365734 &  -.88070922 &   .00000000 \\
 \hline
 1.40247961 &   .91542895 &   .00000000 &  1.40247961 &  -.91542895 &   .00000000 \\
 \hline
 1.30540300 &   .95222319 &   .00000000 &  1.30540300 &  -.95222319 &   .00000000 \\
 \hline
 1.26428480 &   .96444468 &   .00000000 &  1.26428480 &  -.96444468 &   .00000000 \\
 \hline
 1.15140180 &   .98847230 &   .00000000 &  1.15140180 &  -.98847230 &   .00000000 \\
 \hline
  .98655620 &   .99990963 &   .00000000 &   .98655620 &  -.99990963 &   .00000000 \\
 \hline
  .89820891 &   .99480580 &   .00000000 &   .89820891 &  -.99480580 &   .00000000 \\
 \hline
  .78417724 &   .97643256 &   .00000000 &   .78417724 &  -.97643256 &   .00000000 \\
 \hline
  .65832653 &   .93981873 &   .00000000 &   .65832653 &  -.93981873 &   .00000000 \\
 \hline
  .53737749 &   .88655537 &   .00000000 &   .53737749 &  -.88655537 &   .00000000 \\
 \hline
  .34130700 &   .75241181 &   .00000000 &   .34130700 &  -.75241181 &   .00000000 \\
 \hline
  .10715386 &   .45036182 &   .00000000 &   .10715386 &  -.45036182 &   .00000000 \\
 \hline
  .00000000 &   .00000000 &  -1.0000000 & & & \\
 \hline
\end{tabular}
\end{center}
\caption{The eigenvalues of $ D_h $ in a smooth background gauge field
of topological charge Q = 1. The values of other parameters are
$ h_1 = 0.1 $, $ h_2 = 0.2 $, $ A_1^{(0)} = 0.3 $, $ A_2^{(0)} = 0.4 $ and
$ n_1 = n_2 = 1 $. The spectrum shows that $ n_{-} = 1 $ , $ n_{+} = 0 $ and
the Index Theorem and Vanishing Theorem are satisfied exactly. }
\label{table:1}
\end{table}
}

\psfigure 6.0in -0.2in {fig:fig1} {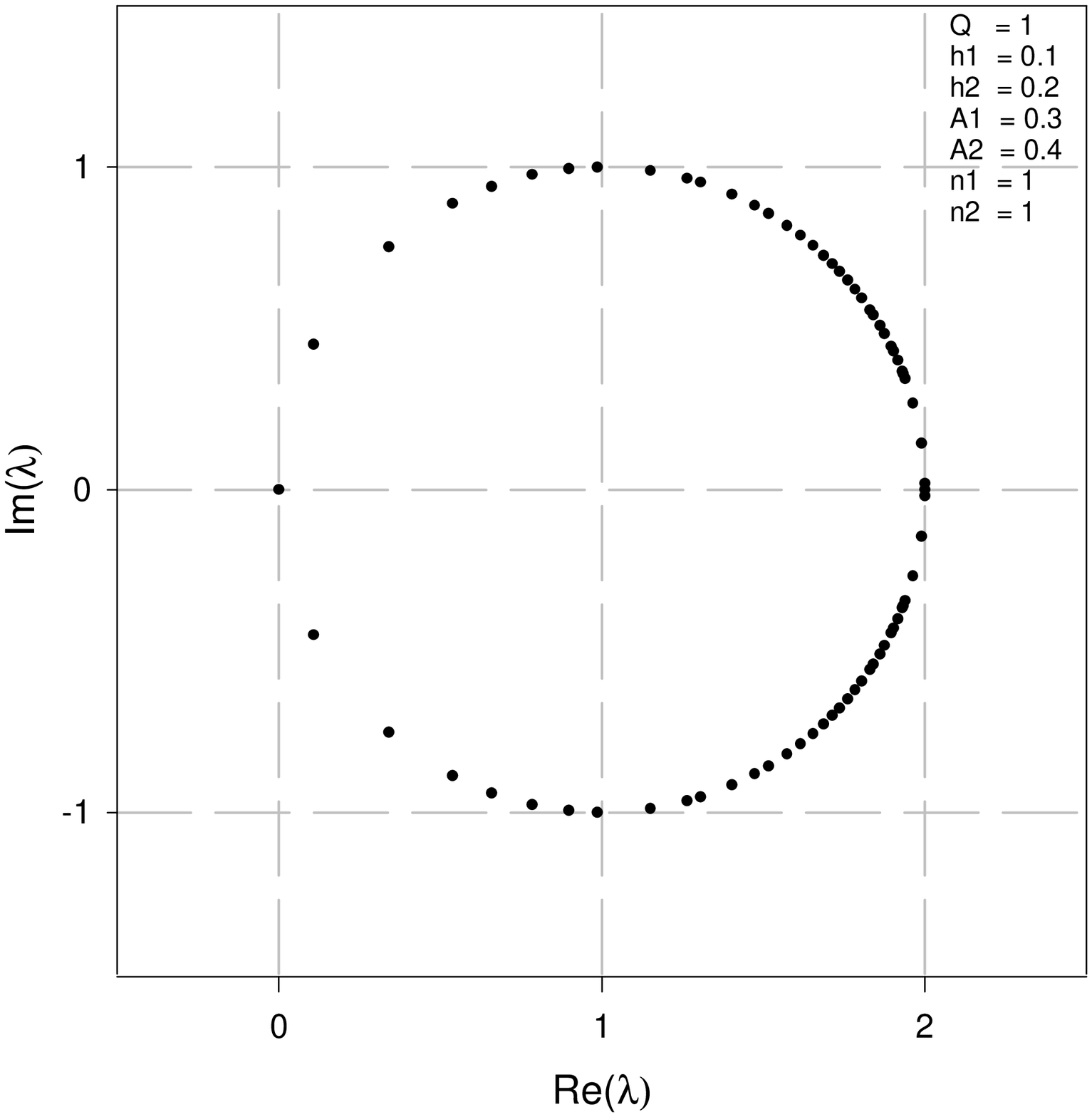} {
The eigenvalues of $ D_h $ in Table 1.
}

{\footnotesize
\begin{table}
\begin{center}
\begin{tabular}{|c|c|c|c|c|c|}
\hline
Re($\lambda$) & Im($\lambda$) & chirality &
Re($\lambda$) & Im($\lambda$) & chirality \\
\hline
\hline
 2.00000000 &   .00000000 &  -1.0000000 & & & \\
 \hline
 2.00000000 &   .00000000 &  -1.0000000 & & & \\
 \hline
 1.99989124 &   .01474800 &   .00000000 &  1.99989124 &  -.01474800 &   .00000000 \\
 \hline
 1.99940122 &   .03460068 &   .00000000 &  1.99940122 &  -.03460068 &   .00000000 \\
 \hline
 1.98160401 &   .19092820 &   .00000000 &  1.98160401 &  -.19092820 &   .00000000 \\
 \hline
 1.94402467 &   .32987486 &   .00000000 &  1.94402467 &  -.32987486 &   .00000000 \\
 \hline
 1.93390001 &   .35753429 &   .00000000 &  1.93390001 &  -.35753429 &   .00000000 \\
 \hline
 1.91210916 &   .40994742 &   .00000000 &  1.91210916 &  -.40994742 &   .00000000 \\
 \hline
 1.90373008 &   .42810273 &   .00000000 &  1.90373008 &  -.42810273 &   .00000000 \\
 \hline
 1.90165787 &   .43245009 &   .00000000 &  1.90165787 &  -.43245009 &   .00000000 \\
 \hline
 1.89104063 &   .45392355 &   .00000000 &  1.89104063 &  -.45392355 &   .00000000 \\
 \hline
 1.87807847 &   .47851667 &   .00000000 &  1.87807847 &  -.47851667 &   .00000000 \\
 \hline
 1.85325494 &   .52149401 &   .00000000 &  1.85325494 &  -.52149401 &   .00000000 \\
 \hline
 1.83821525 &   .54533952 &   .00000000 &  1.83821525 &  -.54533952 &   .00000000 \\
 \hline
 1.81009961 &   .58629226 &   .00000000 &  1.81009961 &  -.58629226 &   .00000000 \\
 \hline
 1.79786888 &   .60283103 &   .00000000 &  1.79786888 &  -.60283103 &   .00000000 \\
 \hline
 1.76948744 &   .63866194 &   .00000000 &  1.76948744 &  -.63866194 &   .00000000 \\
 \hline
 1.74331258 &   .66894425 &   .00000000 &  1.74331258 &  -.66894425 &   .00000000 \\
 \hline
 1.71748554 &   .69657340 &   .00000000 &  1.71748554 &  -.69657340 &   .00000000 \\
 \hline
 1.69419796 &   .71978413 &   .00000000 &  1.69419796 &  -.71978413 &   .00000000 \\
 \hline
 1.66871002 &   .74352331 &   .00000000 &  1.66871002 &  -.74352331 &   .00000000 \\
 \hline
 1.63944000 &   .76884100 &   .00000000 &  1.63944000 &  -.76884100 &   .00000000 \\
 \hline
 1.59262521 &   .80547834 &   .00000000 &  1.59262521 &  -.80547834 &   .00000000 \\
 \hline
 1.54820232 &   .83634575 &   .00000000 &  1.54820232 &  -.83634575 &   .00000000 \\
 \hline
 1.48363430 &   .87527017 &   .00000000 &  1.48363430 &  -.87527017 &   .00000000 \\
 \hline
 1.43095251 &   .90237461 &   .00000000 &  1.43095251 &  -.90237461 &   .00000000 \\
 \hline
 1.39959189 &   .91669315 &   .00000000 &  1.39959189 &  -.91669315 &   .00000000 \\
 \hline
 1.26607166 &   .96395325 &   .00000000 &  1.26607166 &  -.96395325 &   .00000000 \\
 \hline
 1.18891548 &   .98199335 &   .00000000 &  1.18891548 &  -.98199335 &   .00000000 \\
 \hline
 1.06804551 &   .99768222 &   .00000000 &  1.06804551 &  -.99768222 &   .00000000 \\
 \hline
  .90949855 &   .99589632 &   .00000000 &   .90949855 &  -.99589632 &   .00000000 \\
 \hline
  .87904797 &   .99265835 &   .00000000 &   .87904797 &  -.99265835 &   .00000000 \\
 \hline
  .72327655 &   .96094960 &   .00000000 &   .72327655 &  -.96094960 &   .00000000 \\
 \hline
  .56601895 &   .90092200 &   .00000000 &   .56601895 &  -.90092200 &   .00000000 \\
 \hline
  .48688458 &   .85831962 &   .00000000 &   .48688458 &  -.85831962 &   .00000000 \\
 \hline
  .18512327 &   .57963429 &   .00000000 &   .18512327 &  -.57963429 &   .00000000 \\
 \hline
  .00000000 &   .00000000 &   1.0000000 & & & \\
 \hline
  .00000000 &   .00000000 &   1.0000000 & & & \\
 \hline
\end{tabular}
\end{center}
\caption{The eigenvalues of $ D_h $ in a background gauge field
of topological charge Q = -2. The values of other parameters of the
gauge field are the same as in Table 1.
The spectrum shows that $ n_{-} = 0 $, $ n_{+} = 2 $ and
the Index Theorem and Vanishing Theorem are satisfied exactly. }
\label{table:2}
\end{table}
}

\psfigure 6.0in -0.2in {fig:fig2} {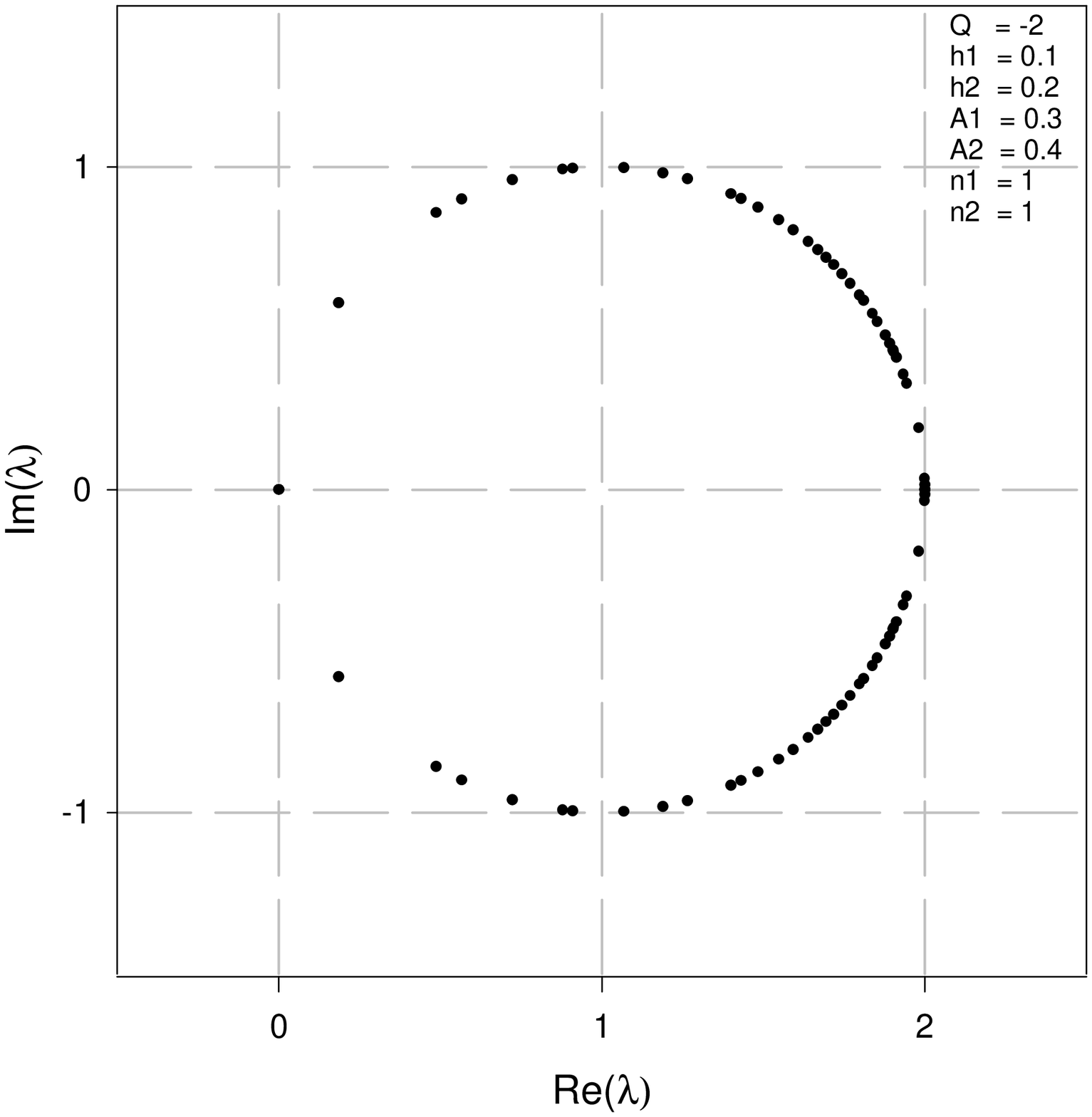} {
The eigenvalues of $ D_h $ in Table 2.
}

{\footnotesize
\begin{table}
\begin{center}
\begin{tabular}{|c|c|c|c|c|}
\hline
   Q  &  $ n_{+} $  &  $ n_{-} $  &  $ n_2^{+} $  &  $ n_2^{-} $  \\
\hline
\hline
  -5  &     5   &     0   &     0   &    5     \\
\hline
  -4  &     4   &     0   &     0   &    4     \\
\hline
  -3  &     3   &     0   &     0   &    3     \\
\hline
  -2  &     2   &     0   &     0   &    2     \\
\hline
  -1  &     1   &     0   &     0   &    1     \\
\hline
   0  &     0   &     0   &     0   &    0     \\
\hline
   1  &     0   &     1   &     1   &    0     \\
\hline
   2  &     0   &     2   &     2   &    0     \\
\hline
   3  &     0   &     3   &     3   &    0     \\
\hline
   4  &     0   &     4   &     4   &    0     \\
\hline
   5  &     0   &     5   &     5   &    0     \\
\hline
\end{tabular}
\end{center}
\caption{The zero modes versus the topological charge.
The Index Theorem and the Vanishing Theorem are satisfied exactly.
Eq.(\ref{eq:chir}) is also satisfied. }
\label{table:3}
\end{table}
}

\section{Fermionic Determinants}

The fermionic determinant $ det(D) $ is proportional to the exponentiation of
the one-loop effective action which is the summation of any number of
external sources interacting with one internal fermion loop.
It is one of the most crucial quantities to be examined in any lattice fermion
formulations. The determinant of $ D_h $ is the product of all its eigenvalues
\bea
det(D_h) = \prod_{s} ( 1 + e^{ \text{i} \theta_s } )
= ( 1 + e^{\text{i} \pi} )^{( n_{+} + n_{-} ) } det_{1}(D_h)
\eea
where $ det_{1}(D_h) $ is equal to the product of all non-zero eigenvalues.
Since the eigenvalues are either real of come in complex conjugate pairs,
$ det(D_h) $ {\em must be real and positive. }
For $ Q = 0 $, then $ n_{+} + n_{-} = 0 $ and $ det(D_h) = det_{1}(D_h) $.
For $ Q \neq 0 $, then $ n_{+} + n_{-} \neq 0 $ and $ det(D_h) = 0 $, but
$ det_{1}(D_h) $ still provides important information about the spectrum.
In continuum, exact solutions of fermionic determinants in the general
background $ U(1) $ gauge fields on a torus was obtained by Sachs and Wipf
\cite{sachs_wipf}. In the following we compute $ det_{1}(D_h) $ for
several different gauge configurations and then compare them with the exact
solutions in continuum. For simplicity, we turn off the harmonic part
( $ h_1 = h_2 = 0.0 $ ) and the local sinusoidal fluctuations
( $ A_1^{(0)} = A_2^{(0)} = 0 $ ) in Eqs. (\ref{eq:A1}) and (\ref{eq:A2}) and
examine the change of $ det_{1}(D_h) $ with respect to the topological charge
$ Q $. For such gauge configurations, the exact solution \cite{sachs_wipf}
is
\bea
det_{1}[ D(Q)] = N \sqrt{ \Bigl( \frac{L_1 L_2 }{2 |Q| } \Bigr )^{|Q|} }
\label{eq:exactD}
\eea
where the normalization constant $ N $ is fixed by
$$
N = \sqrt{ \Bigl( \frac{2}{L_1 L_2 }\Bigr ) }
$$
such that $ det_{1}[D(1)] = 1 $.

In Table 4, the fermionic determinants $ det_1(D_h) $
are listed for $ 8 \times 8 $ and $ 16 \times 16 $ lattices respectively.
They agree with the continuum exact solutions very well for
small $ Q $ but the error goes up to $ 10\% $ for large $ Q \sim 10 $.
For a fixed $ Q $, the error decreases with respect to the increasing
of the size of the lattice.

{\footnotesize
\begin{table}
\begin{center}
\begin{tabular}{|c|c|c|c|c|}
\hline
    & \multicolumn{2}{c|}{ 8 x 8 } & \multicolumn{2}{c|}{ 16 x 16 } \\
\hline
 Q  &   $ det_{1}[D(Q)]_{exact} $  &  $ det_{1}[D_h(Q)] $
    &   $ det_{1}[D(Q)]_{exact} $  &  $ det_{1}[D_h(Q)] $    \\
\hline
\hline
%
  1  &     1.00000  &    1.00000  &   1.00000  &     1.00000   \\
\hline
  2  &     2.82843  &    2.77348  &   5.65685  &     5.66186   \\
\hline
  3  &     6.15840  &    5.96891  &   24.6336  &     24.0615   \\
\hline
  4  &     11.3137  &    10.7157  &   90.5097  &     90.4894   \\
\hline
  5  &     18.3179  &    16.9340  &   293.086  &     286.003   \\
\hline
  6  &     26.8177  &    24.2001  &   858.166  &     822.664   \\
\hline
  7  &     36.1083  &    32.0006  &   2310.93  &     2170.94   \\
\hline
  8  &     45.2548  &    40.2920  &   5792.62  &     5354.28   \\
\hline
  9  &     53.2732  &    45.7353  &   13637.9  &     12309.5   \\
\hline
 10  &     59.3164  &    50.2816  &   30370.0  &     27336.6   \\
\hline
\end{tabular}
\end{center}
\caption{The fermionic determinant versus the topological charge $ Q $.
The normalization constant is chosen such that $ det_{1}[D_h(1)] = 1 $.
The results for the $ 8 \times 8 $ lattice are listed in the second and
the third columns, while for the $ 16 \times 16 $ lattice in the last
two columns. The exact solutions are computed according to
Eq.( \ref{eq:exactD} ). }
\label{table:4}
\end{table}
}

We have also computed the fermionic determinant $ det_{1}[D_h] $ for more
general gauge configurations with non-zero topological charge,
harmonic parts and local parts; as well as in the topologically trivial
sector with only harmonic parts or local parts. They are all in very good
agreement with the continuum exact solutions \cite{sachs_wipf}.

\section{Conclusions and Discussions}

In this paper, we have implemented the square root operator in Neuberger's
proposal \cite{hn97:7} of exactly massless quarks on the lattice by the
recursion formula eq.(\ref{eq:sqrt}). Our numerical tests in two dimensional
background gauge fields assert that $ D_h $ indeed reproduces the exactly
massless fermions on a torus. In particular, for smooth background gauge
fields with non-zero topological charge, the exact zero modes with definite
chirality are reproduced and the Index Theorem is satisfied exactly.

Due to the recent work of Neuberger \cite{hn97:7,hn98:1},
Hasenfratz, Laliena and Niedermayer \cite{ph98:1} and L\"uscher \cite{ml98:2},
we now have a better understanding of the chiral symmetry on the lattice. The
Nielson-Ninomiya theorem \cite{no-go} can be circumvented by replacing the
continuum chiral symmetry $ \{ D, \gamma_5 \} = 0 $ by the Ginsparg-Wilson
relation $ \{ D, \gamma_5 \} = D \gamma_5 D $ on the lattice.
The Ginsparg-Wilson relation gaurantees that the correct continuum anomaly
can be recovered on the lattice. Currently,
there are two classes of solutions satisfying the Ginsparg-Wilson relation.
One of them is the explicit solution $ D_h $ obtained by Neuberger
\cite{hn97:7}, which grew out of the overlap formalism
\cite{rn95,rn93,rd95,yk97:7}. If one attempts to solve the Ginsparg-Wilson
relation by writing $ D = \Id + T $, one must end up with the solution
$ ( \gamma_5 T )^2 = \Id $. Naively one can choose any
lattice operator $ T $ which satisfies this condition
and gives the correct continuum action in the continuum limit. The choice of
$ T = D_w ( D_w^{\dagger} D_w )^{-1/2} $ where $ D_w $ is the standard
Wilson-Dirac action with negative mass term is identical to Neuberger's
overlap solution. We note in passing that for any non-trivial $ T $,
{\em the inverse square root operation in T seems to be inevitable }.
Another class of solutions of Ginsparg-Wilson relation is the fixed point
lattice Dirac operator $ h^{FP} $ \cite{ph98:3} which is the implicit solution
of the non-linear classical saddle-point equations, and is obtained by
recursive iterations. The technical difficulty of that approach is how to
parametrize $ h^{FP} $ such that it is sufficiently precise but the
computational costs are still affordable. It is instructive to
compare our numerical results on zero modes with those obtained by
Farchioni and Laliena \cite{ff98:2} using fixed
point lattice Dirac operator. It is evident that their zero modes are not
exactly zero. The discrepancies are essentially due to the truncation errors
in the parametrization of $ h^{FP} $ and the round-off errors in
the computations performed at each step of the iterations. These errors are
intrinsically mixed together and therefore are very difficult to control.
Furthermore, we suspect that the fermionic determinants $ det_1(h^{FP}) $
of their $ h^{FP} $ would have relatively large errors.
On the other hand, we do not have any serious technical difficulties to
compute $ D_h $ to a very high precision using the recursion formula
eq. (\ref{eq:sqrt}). Therefore we can obtain exact zero modes and fairly
accurate fermionic determinants even on a finite lattice.

It is interesting to note that tracing the Index Theorem on the
lattice has a long history. In 1987, Smit and Vink \cite{smit87} investigated
the spectrum of Wilson and Staggered fermions in topologically
non-trivial background gauge fields in two dimensions. Some remnants of
the index theorem were found but exact zero modes with definite chirality
were not obtained. The Index Theorem on the lattice was also recognized
by Neuberger and Narayanan \cite{rn95, rn97} in the context of overlap
formalism. The index was computed by studying the level crossings in the
spectral flow of $ H ( m ) = \gamma_5 D_w ( m ) $ as a function of $ m $
and was verified numerically to equal to the topological charge
of the background gauge field. When the lattice Dirac fermion operator
$ D_h $ was proposed by Neuberger \cite{hn97:7}, it became clear that the
index computed from level crossings of $ H(m) $ is equal to the index of
$ D_h $.

Although the recursion formula eq.(\ref{eq:sqrt}) is quite effective in
computing $ D_h $ for exactly massless fermions in two dimensions,
however, we expect that it would still be expensive for four dimensional
lattice QCD simulations. An algorithm which invloves at most matrix
multiplications but without matrix inversions at every step of iterations
would be the most desirable. We are now contemplating such possibilities
\cite{bunk98}. There are three emerging investigations that we would like
to carry out in the near future. The first is the spectrum of $ D_h $ in
the four dimensional topologically non-trivial background gauge fields,
which is essentially an extension of the present investigation to four
dimensions \cite{twc98:6}.
The second is the quenched QCD calculations of some chirally
sensitive observables such as the kaon weak matrix elements which were also
calculated by Blum and Soni \cite{bs97} using the Domain Wall quarks
\cite{dbk92,ys93}. Since the Domain Wall quarks in the limit of an infinite
fifth dimension becomes exactly massless and is equivalent to $ D_h $,
it would be interesting to compare future results of using $ D_h $
with those given by Domain Wall quarks with finite fifth dimension
\cite{bs97} and examine whether any improvements can be made by using
$ D_h $ with exactly massless quarks.
The third is to perform dynamical fermion simulations of $ D_h $ in two
dimensions, for example, the Schwinger model. If the results continue to
agree with the continuum exact solutions, we would attempt to confront
one of the most challenging problems in QCD such as the chiral symmetry
breaking by measuring the spectral density of the eigenmodes of $ D_h $
near zero but not exactly zero.

\flushpar {\bf Acknowledgement }

This work is supported by the National Science Council, R.O.C. under the
grant numbers NSC86-2112-M002-017 and NSC87-2112-M002-013. I wish to thank
Herbert Neuberger and Sergei V. Zenkin for comments, remarks and discussions
since the first version of this paper has been posed on the Web. I also wish
to thank Ulli~Wolff for drawing my attentions to ref. \cite{bunk98} and
Philippe~de~Forcrand for suggesting a simple test on dislocations.

\vfill\eject

\vfill\eject

\end{document}